\documentclass[sigconf]{acmart}

\acmDOI{}
\acmISBN{}
\acmArticle{}
\acmPrice{}

\acmConference[recsysXfashion'19]
{Workshop on Recommender Systems in Fashion, 13th ACM Conference on Recommender Systems}
{September 20, 2019}
{Copenhagen, Denmark}
\acmYear{2019}
\copyrightyear{2019}

\usepackage{hyperref}
\hypersetup{
    colorlinks=true,
    linkcolor=blue,
    filecolor=blue,      
    urlcolor=blue,
}
 
\usepackage{tabularx}

\begin{document}

\title{How big can style be? Addressing high dimensionality for recommending with style}


\author{Diogo Goncalves}
\email{diogo.goncalves@farfetch.com}
\affiliation{%
  \institution{Farfetch}
  }

\author{Liwei Liu}
\email{liwei.liu@farfetch.com}
\affiliation{%
  \institution{Farfetch}
  }

\author{Ana Rita Magalh\~aes}
\email{ana.magalhaes@farfetch.com}
\affiliation{%
  \institution{Farfetch}
  }
  
\begin{abstract}
Using embeddings as representations of products is quite commonplace in recommender systems, either by extracting the semantic embeddings of text descriptions, user sessions, collaborative relationships, or product images.
In this paper, we present an approach to extract style embeddings for using in fashion recommender systems, with a special focus on style information such as textures, prints, material, etc. The main issue of using such a type of embeddings is its high dimensionality. So, we propose feature reduction solutions alongside the investigation of its influence in the overall task of recommending products of the same style based on their main image. The feature reduction we propose allows for reducing the embedding vector from 600k features to 512, leading to a memory reduction of 99.91\% without critically compromising the quality of the recommendations.
\end{abstract}



\maketitle

\section{Introduction}
The number and variety of applications for embeddings in the scope of recommender systems is vast. When recommending a fashion product to a user, one can use as a myriad number of ways to create a representation of the product features to enrich the recommendation. It's possible to build embeddings based on the user session data using sequential models like LSTMs. One can also create the semantic representation of a product text description using NLP approaches like word2vec, as well as using Convolutional Neural Networks to extract the visual features present in the product images. Such embeddings can be employed as side information to collaborative models like tensor factorization, use them as inputs to larger models like automated outfit generation \cite{Li2017, HanWJD17}, but also, to the more straightforward task of retrieving similar items based on the extracted embedding space \cite{Gomes2017}.

Deciding whether or not an embedding representation is useful to the recommender system can be a cumbersome task. We need to investigate if the embedding is representing the features that we are looking for to add in the model, if its additional cost is acceptable to the productization process, and, of course, the overall quality improvement expected of the recommender system.

The case-study presented in this paper focuses on visual embeddings. More precisely, on the visual embeddings that can capture the style features\footnote{Throughout this paper, the meanings for style and texture can be considered equivalent. The works from Gatys et al. \cite{Gatys2015, Gatys15A} and the literature on the subject often use the word ``style'' to define the textures of the images being at study. Of course, given the context of luxury fashion, using the word style to describe an image texture and the attributes like print, material and so on, is fertile ground to much controversy and interesting discussions which are out of the scope of this paper.}, or the texture of the product in its detail photos. We have adapted the style representation proposed by Gatys et al. \cite{Gatys15A} to fulfill the need for reducing the size of the Gram matrices, without compromising the information given to the recommendation task. As a recommender system for baseline, we considered a generic content-based filter on the resulting embeddings space to recommend the nearest neighbors to a product query.

Let's consider the task for recommending similar styled products, disregarding their overall shape and category and other generic features. One could use, as product representation, the previous to the top layers of some pre-trained and commonly used convolutional neural networks, such as the GoogleNet Inception V3\cite{HanWJD17}, ResNet50 \cite{Li2017}. Although these embeddings are able to extract the features of the image correctly for the task of finding very similar images, it seems that these type of embeddings tend to focus on generic features of the images, such as the shape and colour of the products being displayed. When the requirement is to represent the style of a product, these embeddings seem to lack the notion of prints and the type of materials being used in the products.

\begin{figure}[h]
  \centering
  \includegraphics[width=\linewidth]{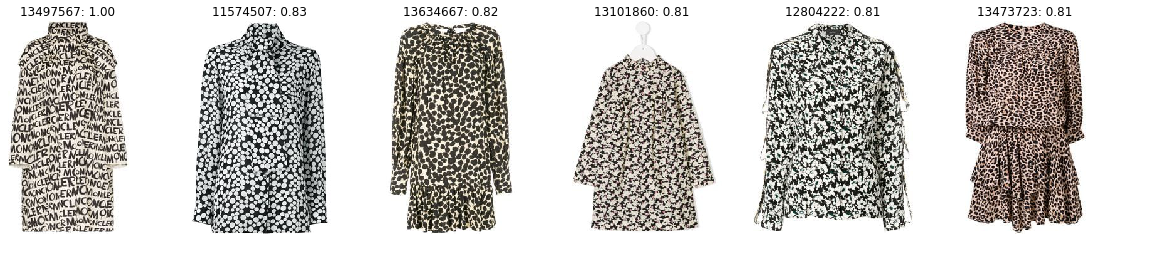}
  \caption{Similar product recommendations to a jacket with graffiti print (the first product being displayed, \url{https://www.farfetch.com/item-13497567.aspx/}).}
  \label{fig:moncler_lookalike}
\end{figure}

Figure \ref{fig:moncler_lookalike} shows a typical response of the similarities computed on the embedding space of the previous to last layer of the ResNet50. The products are part of the Farfetch catalog from where we obtained all the data used in this case-study. The data-set of images comprises the first image (main photo) of a product detail page of the Farfetch.com website that accounts for more than 2 million products of 3000 different designers and high-end luxury brands. These images have the particularity of being taken following a strict set of rules regarding coloring and framing defined by Farfetch fashion authority which is very convenient for case-studies related to computer vision like this one.

If we consider the overall aspect of the query product (the first item on the left of Fig. \ref{fig:moncler_lookalike}), the products being recommended, although being quite similar, they do not capture the graffiti print which is one of the most expressive qualities of the query product. The recommendations seem to focus more on the overall scattering of the colors (black \&white) and its shape (long sleeved top). 

The hypothesis that led to the investigation of using embeddings for style is if the recommender system was using texture representations of the images, then the retrieved products would emphasize more the products' combination of colors, the prints and the texture of materials rather than the overall shape or other features that might be more relevant for the classification task for which the neural networks being used in this study were trained for. Both the ResNet50 and VGG19 CNN architectures used were trained for the ImageNet dataset and we will use the frozen weights to extract image descriptors.

Although using trained neural networks for extracting the embedding representation of product images, there are other approaches in the literature which focus more of the representation of style or textures. We have explored other solutions such as the ones presented in the work from Ustyuzhaninov et al. \cite{UstyuzhaninovBG16} where they employed convolutional filters initialized at random in a shallow neural network. We realized that the results obtained on the recommendations were not worth for pursuing such an approach because the resulting style embeddings were too large for using in production environment and the recommendations were falling too much outside what would be expected for this purpose. The work from Hanbit et al. \cite{Hanbit2017} focused on using a supervised approach to build the style embeddings by training a VGG neural network to map the style based on sets of products borrowing the idea from the Word2Vec method. Although being an interesting approach to map products to defined styles, the need for labeling the products from our dataset into different style classes hindered the implementation of such an approach. Therefore, we decided to conduct this study using as a starting point the proposal from Gatys et al. for texture generation \cite{Gatys15A} where they use a pre-trained CNN on the ImageNet dataset to generate the feature maps from which they extract style representations. 


\section{Methodology}
\subsection{Style embeddings extraction}
Gatys et al. \cite{Gatys2015}, proposed a method for extracting the image texture, or style, by using the feature maps extracted from the VGG19 Neural Network trained on the Imagenet data-set \cite{Simonyan14c}. The main idea is if we compute the feature maps of a certain convolutional layer, we can obtain the texture representation from the correlation between the feature maps in that layer by calculating the Gram matrix of the feature maps. 

To extract the texture representation of a given image $x$, we do a forward pass of that image through the CNN and save the feature map of each filter and of each layer $l$ of the neural network. A layer with $N_l$ filters has $N_l$ feature maps of size $M_l$ if flattened. These feature maps can be represented by as a matrix $F^l \in \mathbb{R}^{N_l \times M_l}$, where $F^l_{jk}$ is the flattened feature map of the $j^{th}$ filter at position $k$ in layer $l$. By computing the Gram matrix $G^l \in \mathbb{R}^{N_l \times N_l}$, where $G^l_{ij}$ as the inner product of the feature map $i$ and $j$ in layer $l$ we get the summary statistic of the feature responses. Hence, we can obtain the Gram matrix $G$ of layer $l$ by the product of the feature map $F^l$ by its transpose $F^{l\top}$:

\begin{equation}
    G^l = F^l \times F^{l\top}
\label{eq:grams}
\end{equation}

From the set of Gram matrices $\{G^1, G^2, ..., G^L\}$  from the layers of interest, the authors built an optimization process to generate images mirroring the representation of texture of the input image $x$. In a paper presented later on \cite{Gatys2015}, they adapted the same process to a multi-objective optimization problem to transfer the style, or texture, from one image to another. This process was named Style Transfer and its most famous result might be the Van Gogh's swirling stars of ``The Starry Night''  transferred to the image of the Neckarfront in T\''ubingen, Germany \cite{Gatys2015}. The architecture proposed by the authors that would lead to better results of style representation and transfer involves extracting and computing the Gram matrices from the feature maps of the first layer of each of the five convolutional blocks of VGG19. Given this solution, the representation of an image is encoded in five square matrices with a number of rows and columns of 64, 128, 256, 515 and 512, respectively. By flattening and concatenating these representations, we would obtain an embedding vector of around 600k elements for each image. Such a vector size could represent a major issue when applying it to a recommender system directly or when adding it as inputs for other models.

The first adaptation of the resulting set of Gram matrices is the selection of the upper triangular part of each of the Gram matrices computed. Since the Gram matrices can be viewed as summary statistics of the feature maps representing the correlation between the filters of each convolutional layer, we expect the matrices to be symmetric and with a negligible diagonal with the correlation between a filter and itself. From Eq. \ref{eq:grams}, we can slice the upper triangle of $G^l$ as the subvector $e^l$.


Let:

$\{a_1, a_2, ..., a_r\}$ be the indices of the r selected rows of the upper triangular part of the Gram matrix and
\\

$\{b_1, b_2, ..., b_s\}$ be the indices of the s selected columns of the upper triangular part (ignoring the diagonal) of the Gram matrix, where $a_1, ..., a_r$ and $b_1, ..., b_s$ are between 2 and $N_l$.
\\

Then the subvector $e^l$ of $G^l$ is denoted as 

\begin{equation}
e^l = G^l [a_1, ..., a_r;b_1, ..., b_s] 
\label{eq:grams_slice}
\end{equation}{}

Each embedding $e^l$ is then normalized by dividing each element by the total size of the feature map $F^l$ for penalizing larger feature maps when concatenating the set of embeddings of the layers in a single vector $e$.

\begin{equation}
    e = [e^1 \cdot \frac{1}{N_1 \cdot M_1}, ..., e^L \cdot \frac{1}{N_L \cdot M_L}]
\end{equation}

for $1, ..., L$ layers of interest.


%

\begin{figure}[h]
\centering
\includegraphics[width=0.5\linewidth]{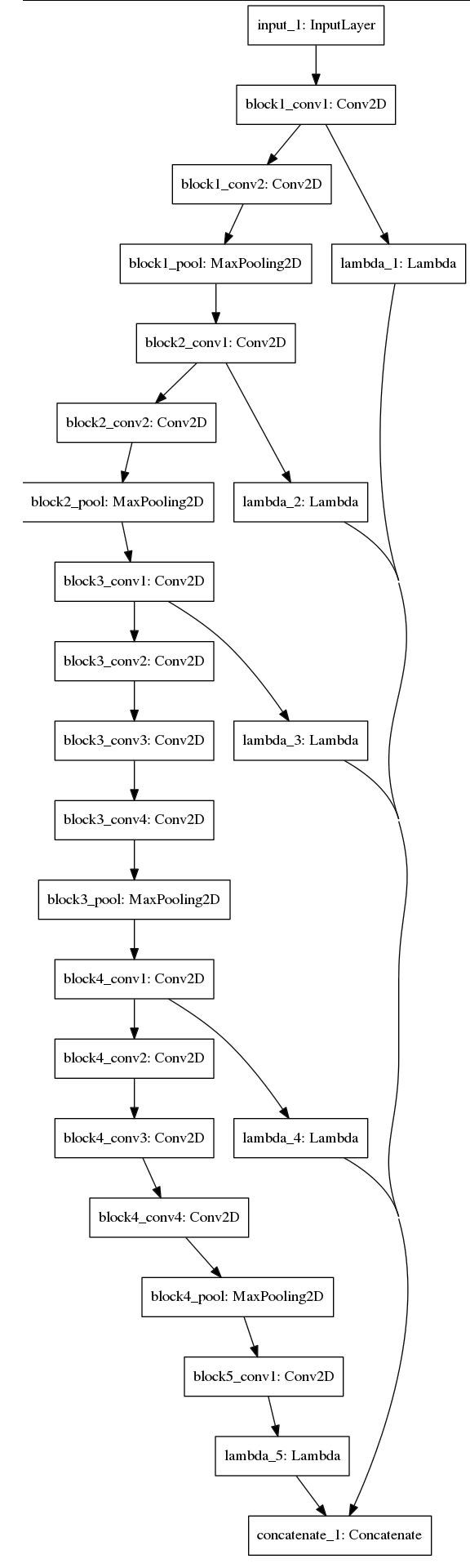}
\caption{VGG19 architecture adapted to style embeddings extraction.}
\label{fig:stylemodel}
\end{figure}

The resulting architecture of the created model derives from the VGG19 model \cite{Simonyan14c} and can be depicted as the model presented in Fig \ref{fig:stylemodel}. The feature maps of the layers of interest are connected to the Lambda layers (see Fig \ref{fig:stylemodel}) which perform the computation of the embeddings of each layer following the methodology described above. 

The selected feature maps are the outputs of the layers of the VGG19 presented in the Table \ref{tab:stylelayers}. Selecting all the layers proposed, and concatenating them after the adaptation we proposed above to reduce the original size, would still produce an embedding vector of 304416 elements. Although being virtually half of the initial size, it still represents a major drawback in using it in other models, or directly in a recommender system for similar products. Therefore, a further study on reducing the dimension is necessary to make it more feasible to production.

\begin{table}[h]
  \caption{Selected layers from VGG19 to construct the style embedding.}
    \label{tab:stylelayers}
    \begin{tabular}{ccc}
    \hline
   \textbf{Layer from VGG19} & \textbf{Gram $G^l$ shape}        & \textbf{Embedding $e^l$ size} \\
   \hline
    block1\_conv1    & $[64 \times 64]$   & 2016                 \\
    block2\_conv1    & $[128 \times 128]$ & 8128                 \\
    block3\_conv1    & $[256 \times 256]$ & 32640                \\
    block4\_conv1    & $[512 \times 512]$ & 130816               \\
    block5\_conv1    & $[512 \times 512]$ & 130816             \\
    \hline
    \end{tabular}
\end{table}

The embeddings extracted from the ResNet50 previous to the top layer have 2048 features which is significantly smaller than the embeddings extracted following the approach proposed above. The utility of adding such embeddings to the recommendations is assessed in the results section \ref{sec:results}, by comparing the recommendations of both approaches, ResNet50 embedding versus style embeddings.

\subsection{Dimensionality reduction by sampling the feature maps}
The idea behind this method is to randomly sample a percentage of the feature maps of each layer of interest (Fig. \ref{fig:stylemodel}) to reduce the overall dimension of the embedding vector. The assumption is that some of the feature maps might carry enough information about the texture captured at that layer. Hence, we need to do this operation before the computation of the Gram matrices to guarantee that the result of each Gram is the correlation between selected filters response. Taking the matrix of the feature maps $F^l$ for layer $l$, we can obtain a submatrix $F^{l*}$ by selecting a specific number of rows (feature maps) at random.

\begin{equation}
    F^{l*} = F^{l}[r_1, r_2, ..., r_N]
\end{equation}

where 
$\{r_1, r_2, ..., r_s\}$ are the uniformly sampled indices $r$ of the $p$ selected rows. $p$ is the nearest integer obtained by the proportion $\alpha$ of features maps to be selected:

\begin{equation}
    p = \lfloor \alpha N_l \rceil
\end{equation}

where $\alpha \in [0, 1]$.

As a result, rather than having a feature map result of $[N_l \times M_l]$, we obtain a feature map of  $[\alpha N_l \times M_l]$, which will result in a square gram matrix of shape $[\alpha N_l \times \alpha N_l]$ and, consequently, on an embedding vector of size $\frac{1}{2}\alpha N_l (\alpha N_l-1)$ . For the example of the convolutional layer ``block1\_conv1'' of the model represented in Fig. \ref{fig:stylemodel} would naturally result in an embedding vector of size 2016 ($\frac{1}{2}N_l (N_l-1)$). When sampling 50\% of the feature maps from that layer, we get an embedding vector of 496 features which results in a dimensional reduction of 70\%. If we consider the convolutional layer 5 ``block5\_conv1'', we reduce the embedding dimension to 32640 rather than the original size of 130816 that the complete feature maps would return. We conducted a grid search for $\alpha$ variations and compared the recommendations obtained after sampling the feature maps against the baseline model described above in Fig \ref{fig:stylemodel} with the metrics of Precision@100 and nDCG@100.

\subsection{Dimensionality reduction with Truncated SVD}
We tried to implement a PCA via an encoder-decoder approach directly in the baseline model of Fig \ref{fig:stylemodel} but it was not memory efficient. Therefore, we employed the Truncated SVD which was efficient and was providing good results. We conducted a grid search for the resulting embedding sizes of \{2048, 1024, 512, 128, 64\} and compared the performance against the baseline model with the metrics of Precision@100 and nDCG@100.

\section{Results}
\label{sec:results}
\subsection{Similar style recommendations}
The first issue to address is if the new embeddings, using the baseline model Fig. \ref{fig:stylemodel}, are recommending products with a higher emphasis on the style, textures, prints, materials, etc, than the recommender using the embeddings of the ResNet50 CNN.

We sampled around 5K products and generated recommendations for a smaller subset of products for the different type of embeddings and products to then ask a group of fashion experts at Farfetch, using an internal tool, to evaluate all the top 10 recommendations from each of the embeddings. The goal was to see which would be better at recognizing the elements of style, such as texture of the product, type of the print, type of material and colouring.

To reduce the number of recommendations that would need to evaluated, we sampled around 5K products by looking at search results from the internal search engine. On top of them, we also randomly selected some products to introduce some randomness to the data-set. We chose the following search terms to guarantee a pool of products with distinctive features:

\begin{itemize}
    \item ``denim'',
    \item ``oversized'',
    \item ``glitter'',
    \item ``motif'',
    \item ``striped'',
    \item ``checked'',
    \item ``lace'',
    \item ``leather'',
    \item ``sandal'',
    \item ``bucket'',
    \item ``graffiti'',
    \item ``floral print'',
    \item ``snake print'',
    \item ``zebra print'',
    \item ``tiger print'',
    \item ``ruffled blouse'',
    \item ``chain'',
    \item ``burberry checked print'',
    \item ``louis vuitton''. 
\end{itemize}

The overall comparison regarding Precision@10 between the recommendations using the style embeddings and the embeddings extracted by the ResNet50 can be seen on the Table \ref{tab:styleresults}. The adapted version of VGG19 following the method we propose is performing slightly better at identifying products sharing style features than the recommender using the ResNet50 features. 

\begin{table}[h]
  \caption{Voting results for assessing if the recommendations are capturing style features.}
    \label{tab:styleresults}
\begin{tabular}{lrrr}
\toprule
\textbf{Embeddings model} &  \textbf{Precision@10} &      \textbf{SD} &    \textbf{Pop. size} \\
\midrule
VGG19 adapted & 0.38 &  0.21 &  177 \\
ResNet50 & 0.35 &  0.18 &  177 \\
\bottomrule
\end{tabular}
\end{table}

Although not presenting very significant results, we hypothesize that the proposed approach is able to capture the features that could describe the product in terms of style.

\begin{figure*}
   \centering
\begin{tabular}{rl}
Style embedding &
\includegraphics[scale=0.33]{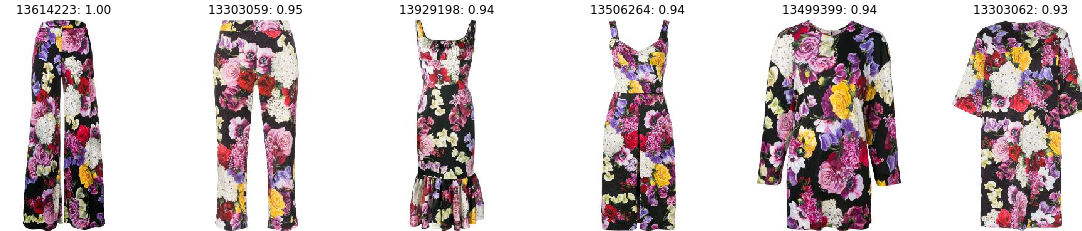}
\\
ResNet50&
\hspace*{0.2cm}\includegraphics[scale=0.33]{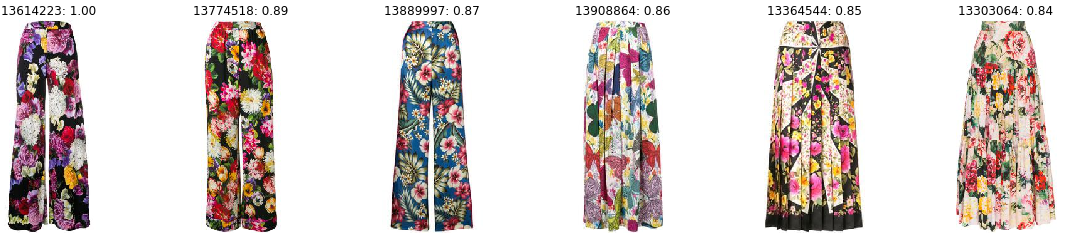}\\
\end{tabular}

  \caption{Floral print recommendations comparison from the product query displayed first from the left. Top row: Adapted version of VGG19 embeddings. Bottom row: ResNet50 embeddings}
  \label{fig:style_lookalike_floral}
\end{figure*}

\begin{figure*}
   \centering
\begin{tabular}{rl}
Style embedding &
\includegraphics[scale=0.33]{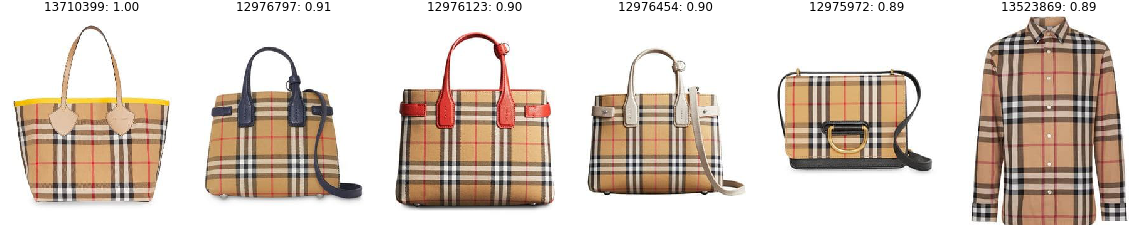}
\\
ResNet50&
\hspace*{0.0cm}\includegraphics[scale=0.33]{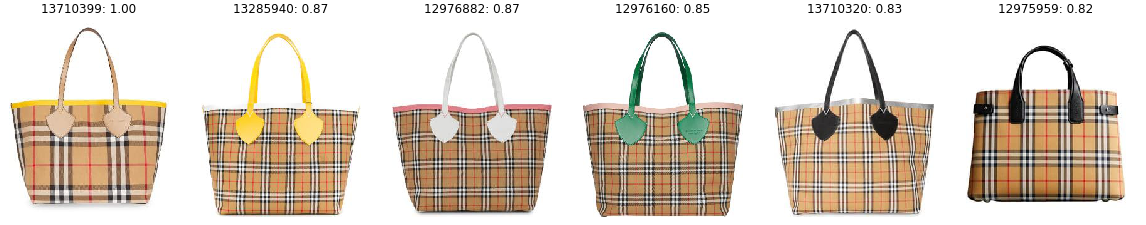}\\
\end{tabular}

  \caption{Burberry checked print recommendations comparison from the product query displayed first from the left. Top row: Adapted version of VGG19 embeddings. Bottom row: ResNet50 embeddings}
  \label{fig:style_lookalike_burberry}
\end{figure*}

\begin{figure*}
   \centering
\begin{tabular}{rl}
Style embedding &
\includegraphics[scale=0.33]{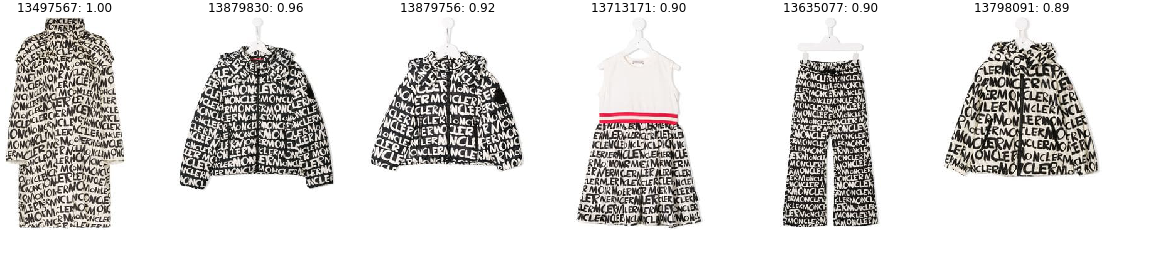}
\\
ResNet50&
\hspace*{-0.2cm}\includegraphics[scale=0.33]{moncler_lookalike.png}\\
\end{tabular}

  \caption{Moncler graffiti print recommendations comparison from the product query displayed first from the left. Top row: Adapted version of VGG19 embeddings. Bottom row: ResNet50 embeddings}
  \label{fig:style_lookalike_moncler}
\end{figure*}

By doing a deep-dive on the resulting recommendations (Figs. \ref{fig:style_lookalike_floral},  \ref{fig:style_lookalike_burberry},  \ref{fig:style_lookalike_moncler}), we can observe that the distinctive features of prints, textures and colouring are being given more weight than the overall shape of the product when using the style embeddings versus the ResNet50 embeddings. The selected products capture distinctive prints and styles that highlight the need for different embeddings to describe the images of the products.

The floral print of the product query of Fig. \ref{fig:style_lookalike_floral} has a very similar design to the recommendations using style embeddings whereas with the embeddings of ResNet50, the recommendations seem to be more focused on the pants category. If we look closer, we see the same shape of flowers being recommended in the style embeddings version.

The recommendations of style for the Burberry print of Fig. \ref{fig:style_lookalike_burberry} show a very similar checked print to the query product by being larger than the recommendations of the ResNet50 embeddings which emphasized the model of the bag, disregarding the print detail. As we can see, the style embeddings were able to find similar products across different product categories.

Lastly, the product we started with in the introduction session gets recommendations showing the exact same graffiti print as the query product, contrary to the prints of a black and white dispersion of the products being shown by the counterpart embeddings of ResNet50. 

These results suggest the effectiveness of the style embeddings as being able to capture the elements of style of the products. Now, we need to address the dimensionality reduction to be able to use this information in other models and in production environments.

\subsection{Dimensionality reduction by sampling the feature maps}

We computed the similarities between all the products sampled and obtained the top 100 nearest neighbors to each product according to the cosine similarity on the embedding space of the baseline model. 

To understand the loss of quality by the process of randomly slicing the feature maps, we computed the top 100 recommendations to the new embeddings obtained in the smaller feature maps and compared the recommendations against the baseline model which has 100\% of the feature maps ($\alpha = 1$). 

Table \ref{tab:sampleresults} shows the loss in precision and nDCG due to the removal of information from the feature maps. As we can see, the loss of precision is more severe than the nDCG. This observation reflects that, although the number of products being matched in the top 100 recommendations are different, the head of the list of recommended products are following a similar order, leaving the tail of the recommendations to be ranked differently. Therefore, this approach might be useful for dealing with high dimensionality in such scenarios. 

\begin{table}[h]
  \caption{Precision and nDCG results for dimensionality reduction by sampling the feature maps.}
    \label{tab:sampleresults}
\begin{tabular}{rrrr}
\toprule
\textbf{Feature maps }& \textbf{Embedding $e^l$} &  \textbf{Precision} & \textbf{nDCG} \\
\textbf{proportion, $\alpha$}& \textbf{size} &\textbf{@100} & \textbf{@100}\\
\midrule
       100\% &           304416 &   1    &  1 \\
        70\% &           148732 &   0.95 &  0.98 \\
        50\% &            75920 &   0.91 &  0.97 \\
        30\% &            27362 &   0.87 &  0.96 \\
        10\% &             2968 &   0.71 &  0.90 \\
\bottomrule
\end{tabular}
\end{table}

The fact that removing information is not hindering the recommendations severely might indicate that some of the filters are mapping irrelevant information about style or textures. Hence, for future works, we plan to pinpoint which filters are irrelevant for this task and remove them directly from the original VGG19 before constructing our proposed model.

\subsection{Dimensionality reduction by Truncated SVD}
The results of dimensionality reduction using Truncated SVD shown on Table \ref{tab:svdresults} reflect an almost perfect fit to the recommendations being computed by the baseline model with the 300k if we consider a reduction to 2048 components.

\begin{table}[h]
  \caption{Precision and nDCG results for dimensionality reduction by Truncated SVD.}
    \label{tab:svdresults}
\begin{tabular}{rrr}
\toprule
\textbf{Embedding $e^l$} &  \textbf{Precision} & \textbf{nDCG} \\
 \textbf{size} &\textbf{@100} & \textbf{@100} \\
\midrule
            2048 &   0.992 &  0.997 \\
            1024 &   0.982 &  0.992 \\
             512 &   0.966 &  0.985 \\
             128 &   0.913 &  0.961 \\
              64 &   0.865 &  0.937 \\
\bottomrule
\end{tabular}
\end{table}

Even the most aggressive reduction, using the 64 features embedding, have an equivalent performance to the sampling of 30\% of the feature maps of the previous section. 

The embedding vector of size 512 seems to be the most efficient regarding size, precision, and computational power needed for the reduction processing. Moreover, if the productization of truncated SVD is not an issue, then the previous method of sampling might be unnecessary. However, there might be some technical challenges to overcome if the Truncated SVD needs to run for a pool of 2 million products. In that case, alternatives such as the Incremental PCA, which performs the update of the components in batch, might be better suited for performance regarding computational cost.

\section{Conclusion and future work}
In this paper we presented a way of extracting and using style embeddings in an efficient approach which allows for using it as side information for finding similar styled products or to complement more complex recommender systems without compromising its feasibility in production environments.

The results have shown that the embeddings used for fashion purposes might require some more in depth analysis to guarantee that they are capturing what is intended. In our example, we compare the embeddings of ResNet50 with our derivation from VGG19, and it was found that features that might be more useful to capture the style of a product were better captured by our approach. Both NNs used in this paper were trained with the same purpose of classifying 1000 labels from the ImageNet dataset. However, using the embeddings from the Gram matrices proposed by Gatys et al. \cite{Gatys2015} might promote a richer type of information when it comes to describe the images of products to build fashion recommender systems.

The reduction approaches have proven to be effective without compromising the quality of the recommendations. Moreover, we saw that there might be quite a significant amount of non useful information being captured by the inner convolutional layers of the CNN, which leads us to investigate the subject further and pin-point the filters which might be irrelevant for this task.

As future work, we plan to incorporate the style information in our Collaborative-based models and use them in a Siamese Neural Network for generating automated outfits with a better sense of style.

\bibliographystyle{ACM-Reference-Format}
\bibliography{recsys19farfetch}

\end{document}